\documentstyle[floats,aps,prl,epsfig,twocolumn]{revtex}

\begin{document}

\wideabs{

\title { 
Theory of Distinct Crystal Structures of Polymerized Fullerides
AC$_{60}$, A=K, Rb, Cs: the Specific Role of Alkalis
} 
 
\author {K.H.~Michel$^1$ and  A.V.~Nikolaev$^{1,2}$ \\} 

\address{
$^1$Department of Physics, University of Antwerp, UIA, 
2610 Antwerpen, Belgium\\
$^2$Institute of Physical Chemistry of RAS, 
Leninskii prospect 31, 117915, Moscow, Russia 
} 
 
\date{\today} 
 
\maketitle 
 
\begin{abstract} 
The polymer phases of AC$_{60}$ form distinct crystal structures
characterized by the mutual orientations of the $(C_{60}^-)_n$ chains.
We show that the direct electric quadrupole interaction between
chains always favors the orthorhombic structure $Pmnn$ with
alternating chain orientations.
However the specific quadrupolar polarizability of the alkali metal ions
leads to an indirect interchain coupling which favors the monoclinic
structure $I2/m$ with equal chain orientations.
The competition between direct and indirect interactions explains
the structural difference between KC$_{60}$ and RbC$_{60}$,
CsC$_{60}$.
\end{abstract} 
 

}

\narrowtext 
 
Alkali metal doped C$_{60}$ (A$_x$C$_{60}$), A=K, Rb, Cs,
form stable crystalline phases (fullerides) with a broad
range of physical and chemical properties comprising
superconductors and polymer phases.
For a review, see \cite{Dre,Kuz}.
In particular the $x=1$ compounds \cite{Zhu} exhibit plastic
crystalline phases with cubic rocksalt structure at high
temperature ($T \ge 350$~K) and polymeric phases \cite{Pek,Cha,Ste}
of reduced symmetry at lower $T$.
In the latter the C$_{60}$ molecules are linked through
a [2+2] cycloaddition \cite{Ste}, a mechanism originally
proposed for photo-induced polymerization \cite{Rao}
in pristine C$_{60}$. From X-ray powder diffraction \cite{Ste}
it was concluded that the crystal structure of both KC$_{60}$
and RbC$_{60}$ was orthorhombic (space group $Pmnn$).
Polymerization occurs along the orthorhombic $\vec{a}$ axis
(the former cubic [110] direction), where the orientation of the
polymer chain is characterized by the angle $\mu$ of the planes
of cycloaddition with the $\vec{c}$ axis.
In the $Pmnn$ structure (Fig.~1a), these orientations are
alternatively $\mu$ and $-\mu$, $|\mu| \approx 45^0 \pm 5^0$.
Notwithstanding this apparent structural similarity,
the electronic and magnetic properties of KC$_{60}$
on one hand and RbC$_{60}$, CsC$_{60}$ on the other hand,
were found to be very different \cite{Kuz}.
ESR and optical conductivity data
\cite{Cha,Bom} show that RbC$_{60}$ and CsC$_{60}$ exhibit a transition
from a quasi-one dimensional metal to an insulating magnetic state near 50~K,
while KC$_{60}$ stays metallic and nonmagnetic at low $T$.
NMR spectra also showed marked differences between KC$_{60}$
and Rb-, Cs- polymers \cite{All}.
The contradiction between similar crystalline structures and
different electromagnetic properties was resolved by single crystal
X-ray diffraction and diffuse scattering studies \cite{Lau}.
Indeed the polymer phases of KC$_{60}$ and RbC$_{60}$ are
different. While the space group $Pmnn$ is confirmed for
KC$_{60}$, it is found that RbC$_{60}$ is body centered
monoclinic, with space group $I2/m$. In the latter structure,
the polymer chains have the same orientation $\mu$ (Fig.~1b).
Electronic band structure calculations for the $I2/m$ structure
have shown the importance of transverse interchain
coupling for RbC$_{60}$ \cite{Erw}.
Recently \cite{Rou} high-resolution synchrotron powder 
diffraction results demonstrated that CsC$_{60}$ has the same
structure as RbC$_{60}$. 
%
\begin{figure} 
\vspace{-1.1cm}
\centerline{
\epsfig{file=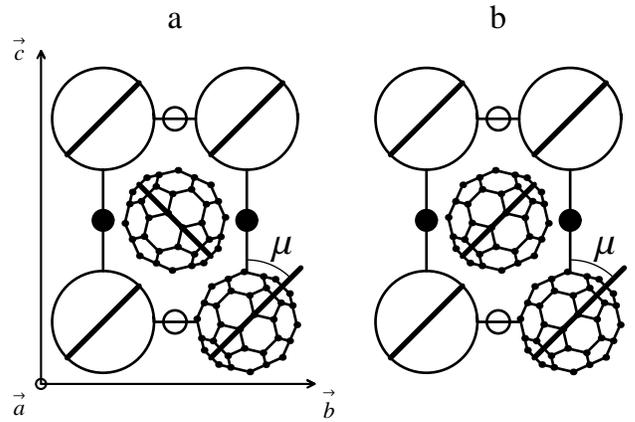,width=0.54\textwidth}
} 
\vspace{-7.1cm}
\caption{
Crystal structures projected onto the
crystallographic $(\vec{b},\vec{c})$ plane:
(a) $Pmnn$, (b) $I2/m$. The thick bars represent the projection
of the cycloaddition planes.
Polymerization occurs along $\vec{a}$. The alkalis located in 
$(\vec{b},\vec{c})$ planes and at $\pm a/2$ are denoted by full $(+)$
and empty $(-)$ circles.
} 
\label{fig1} 
\end{figure} 

In this paper we study the mechanism which leads to the distinct
polymer phases and we demonstrate the active role of the 
distinctive quadrupolar electronic polarizability of the alkali 
ions (cations). We start from the high temperature orientationally 
disordered cubic phase (space group $Fm{\bar 3}m$) and describe
the formation of the polymer phases as a scenario
with several steps:
i) the charge transfer of one electron from the alkali atom
to the C$_{60}$ molecule leads to an occupation of the lowest
unoccupied molecular orbital levels which are of $t_{1u}$
symmetry. Thereby the crystal field of the C$_{60}^-$ ion
acquires an electronic component \cite{Nik} which favors
a same orientation of neighboring molecules
along [110] such that the stereospecific cycloaddition occurs.
ii) The cycloaddition between neighboring molecules then
acts as a negative internal stress (chemical pressure)
along [110]. 
iii) The mutual orientation of neighboring C$_{60}^-$ chains, which
distinguishes between the orthorhombic and monoclinic structures, 
should depend on the intercalated alkali ions.

We first show that the orthorhombic lattice is the result of
the cycloaddition.
Using concepts of the theory of elasticity
\cite{Lan}, we find that the cubic crystal is deformed into 
an orthorhombic one (point group $D_{2h}$).
Taking the cubic [110], [1${\bar 1}$0] and [001] as new
$x$, $y$ and $z$ axes 
(orthorhombic $\vec{a}$, $\vec{b}$ and $\vec{c}$)
respectively, we find the deformations 
\begin{mathletters} 
\begin{eqnarray} 
& & \epsilon_{xx}=\frac{K}{d c_{44}} [c_{11}(c_{11}+c_{12}+2c_{44})-2c_{12}^2],
  \label{1a} \\
& & \epsilon_{yy}=-\frac{K}{d c_{44}} [c_{11}(c_{11}+c_{12}-2c_{44})-2c_{12}^2],
  \label{1b} \\
& & \epsilon_{zz}=-K c_{12}/d , \label{1c}
\end{eqnarray}
\end{mathletters}
where $c_{ij}$ are the cubic elastic constants, 
$d=c_{11}(c_{11}+c_{12})-2c_{12}^2$ and where $K<0$ is the uniaxial stress.
Obviously $\epsilon_{xx}<0$, $\epsilon_{yy}>0$ and $\epsilon_{zz}>0$
which corresponds to contraction along $\vec{a}$ and elongations
along $\vec{b}$ and $\vec{c}$.
We next investigate the orientation of the polymer chains.
Following the experimental breakthrough \cite{Lau}, preliminary
calculations of interchain energies have been performed \cite{Lau2}.
The interaction energy is highly sensitive to the lattice constants 
and a plausible scenario is that
C$_{60}$-C$_{60}$ interchain distances impose different relative chain
orientations.
However the last argument is at variance with single crystal X-ray
diffraction results of pressure polymerized C$_{60}$ \cite{Mor}.
There the space group is $Pmnn$, $i.e.$
isostructural with polymerized KC$_{60}$.
However the orthorhombic cell volume of polymerized C$_{60}$
and the distance between a corner and the center of the cell,
1326 {\AA}$^3$ and 9.956 {\AA}, respectively, are closer
to the corresponding values in RbC$_{60}$, 1314.9 {\AA}$^3$
and 9.852 {\AA}, than to those of KC$_{60}$, 1298 {\AA}$^3$ and 
9.836 {\AA}.
We conclude that the alkalis must play a more specific role
in triggering the structural difference between KC$_{60}$
and RbC$_{60}$, CsC$_{60}$. From the theory of bilinear
rotation-translation (RT) coupling between molecular rotations 
and lattice displacements of the counterions 
which plays an essential role in determining the elastic 
properties of ionic molecular crystals \cite{Lyn}, one
finds an effective rotation-rotation (RR) interaction that 
competes with the direct intermolecular RR interaction.
However the lattice mediated RR interaction is
found to be independent of the alkali mass and hence no 
distinction on basis of the different masses is possible.
A further distinctive property of the alkali metal ions is 
the dipolar electronic polarizability, 
with values 0.9, 1.7 and 2.5 {\AA}$^3$ for
K$^+$, Rb$^+$ and Cs$^+$, respectively \cite{Ash}.
We expect that also the quadrupolar polarizability is larger
for the heavy alkali metal ions than for K$^+$.
In the following we will show that the electric quadrupole
interaction between C$_{60}^-$ chains and the alkali ions leads to
an effective orientational interaction between the  C$_{60}^-$
chains.

We will study first the electric multipole interaction
between polymer chains.
Polymerization reduces the symmetry, the C$_{60}^-$ chain is
composed of units with $D_{2h}$ symmetry.
We have used a tight-binding model to study the electronic charge
distribution on the C$_{60}^-$ units in the chain \cite{Nik2}.
The charge is mainly concentrated in the equatorial region of
C$_{60}^-$, in agreement with recent NMR results \cite{Swi}.
We find that only even $l$ multipoles are allowed, in particular
each C$_{60}^-$ unit has an electric quadrupole.
In the following we adopt a simple model of charge
distribution. Using the labeling of C atoms of \cite{Swi},
we locate a charge of $-0.15$ (units $|e|=1$) on each bond
C15-C16.
These charges are fixed at a distance $d=3.52$ {\AA} from the
center of the C$_{60}^-$ ball.
Such a charge distribution is sufficient to obtain a quadrupole
(see also Eq.~(\ref{2})), the accompanying monopole is irrelevant.
The chains are taken as rigid units with sole degree of freedom
the rotation angle $\mu$ about the axis $\vec{a}$.
Hence we can treat the three dimensional crystal in the polymer
phase as a two dimensional problem in the $(\vec{b} \vec{c})$
plane. 
Here we consider one chain per unit cell
with basis vectors $\vec{r}_1=(\vec{c}/2)+(\vec{b}/2)$
and $\vec{r}_2=(\vec{c}/2)-(\vec{b}/2)$.
The chains are labeled by a two dimensional array
$\vec{n}=(n_1,n_2)$ and lattice vectors are given by
$\vec{X}(\vec{n})=n_1 \vec{r}_1+n_2 \vec{r}_2$, where
$n_1$, $n_2$ are integers.
The Coulomb interaction between chains depends on their
mutual orientation.
We  introduce symmetry adapted rotator functions (SAF's)
$S_l(\vec{n})=\sin(l \mu(\vec{n}))$.
Symmetry of the chain for $\mu \rightarrow \mu+\pi$ implies
$l=2$, 4, ...~. 
These functions are ungerade in $\mu$ and $S_2$ is maximum
for $\mu=45^0$.
Expanding the Coulomb interaction in terms of
SAF's we consider in lowest order of $l$
the quadrupole-quadrupole
interaction
\begin{eqnarray} 
 H_{SS}=\frac{1}{2}  \sum_{\vec{n}\, \vec{n}'}
 (J_a(\vec{n},\vec{n}')+
 J_b(\vec{n},\vec{n}') )\, S_2(\vec{n})\, S_2(\vec{n}') . \label{2}
\end{eqnarray}
More involved molecular charge distributions would lead
to higher ($l \ge 4$) multipoles \cite{Nik}, but their interactions are
negligible.
We observe that nearest neighbor chains are shifted by a translation
$a/2$ along the $\vec{a}$ axis with respect to each other.
For a C$_{60}^-$ molecule at $\vec{n}$ as origin, 
$J_a$ takes into account the interaction
with the two molecules at $\pm a/2$ on the neighboring four
chains $X(\vec{n}')=\pm \vec{r}_1$ and $X(\vec{n}')=\pm \vec{r}_2$, 
$J_b$ describes the interaction
with one molecule on the chains $X(\vec{n}')=\pm (\vec{r}_1-\vec{r}_2)$.
In two dimensional Fourier space we obtain
\begin{mathletters} 
\begin{eqnarray} 
 & &H_{SS}= \frac{1}{2} \sum_{\vec{q}}
 J(\vec{q})\, S_2(\vec{q})\, S_2(-\vec{q}) , \label{3a}  \\
 & &J(\vec{q})=8 J_a\, \cos \left( \frac{q_y b}{2} \right) 
 \cos \left( \frac{q_z c}{2} \right)
 +2 J_b \cos(q_y b) , \label{3b}
\end{eqnarray}
\end{mathletters}
where $q_y$ and $q_z$ are components along the original orthorhombic
axes $\vec{b}$ and $\vec{c}$.
With the simple model of charge distribution for the C$_{60}^-$
units and the orthorhombic lattice constants of Table~I,
we have calculated the quadrupolar interaction energies $J_a$
and $J_b$ quoted in Table I.
%
%
\begin{table} 
\caption{ 
Lattice constants $a_c$ refer to cubic,
$a_o$, $b_o$ and $c_o$ to simple orthorhombic lattice cells,
length in units {\AA} (Angstr\"{o}m). Interaction energies
$J_a$, $J_b$, $\lambda_b$, $\lambda_c$, $g_A$ measured in units K (Kelvin).
Quadrupolar radii $d_A$ in units {\AA}. Ratios of $g_A$'s are equal
to the corresponding ratios of $1/d_A$'s.
\label{table1}     } 
 \begin{tabular}{c | c | c c c | c c c c c c} 
 &  $a_c$ & $a_o$ & $b_o$ & $c_o$ & $J_a$ & $J_b$ & $d_A$ & 
                           $\lambda_b$ & $\lambda_c$ & $g_A$\\ 
\tableline 
 K & 14.06 & 9.11 & 9.95 & 14.32 & 16.7 & -86.2 & 1.47 & 14.5 & 80.0 & 173 \\ 
 Rb& 14.08 & 9.14 & 10.11& 14.23 & 17.4 & -79.0 & 1.82 & 20.7 & 130.5 & 140 \\
 Cs& 14.13 & 9.10 & 10.22& 14.17 & 18.2 & -74.5 & 1.87 & 22.3 & 141.8 & 136 
 
 \end{tabular} 
\end{table} 
From Eq.~(\ref{3b}) we find that $J(\vec{q})$ is negative and
maximum in absolute value at the boundary of the Brillouin zone (BZ).
For all three compounds we have 
$|J(\vec{q}_Z)|>|J(\vec{q}_{\Gamma})|$, where
$\vec{q}_Z=(0,2\pi/c)$ is a BZ boundary vector and
$\vec{q}_{\Gamma}=(0,0)$ is the BZ center.
This result is independent of the strength of the molecular quadrupoles
but is a consequence of the orthorhombic lattice.
The dominance of $J(\vec{q}_Z)$ leads to
a condensation of $S_2(\vec{q})$
for $\vec{q}=\vec{q}_Z$: $\langle S_2(\vec{q}_Z) \rangle = \eta/\sqrt{N} $.
Here $\eta$ is the order parameter amplitude and $N$ is
the number of chains ($i.e.$ lattice points in the $(\vec{b} \vec{c})$ plane).
Condensation at $\vec{q}_Z$ implies that the chains in a same basal plane
$(\vec{a} \vec{b})$ of the orthorhombic lattice 
all have the same orientation, but the
orientation alternates in neighboring planes at distance $c/2$.
This is the ``antiferrorotational" structure $Pmnn$, Fig.~1a.
We find that the quadrupolar interaction between C$_{60}$ chains
leads to $Pmnn$ for KC$_{60}$, RbC$_{60}$, CsC$_{60}$,
irrespective of the different orthorhombic lattice constants.
For RbC$_{60}$ and CsC$_{60}$ the experimental result is
$I2/m$ \cite{Lau,Rou}.
We now include the role of the alkali ions with their distinctive
polarizability.
It is known from work on the ammonium halides \cite{Hul}
that the indirect interaction of two NH$_4^+$ tetrahedra via
the polarizable halide ions plays an essential role  
in determining 
the various crystalline phases of the ammonium halides
NH$_4$X, X=Cl, Br, I \cite{Par}.
However in the present problem, by symmetry the
 dipolar polarizability of the alkali metal ions
is irrelevant and we have to resort to the quadrupolar
polarizability.
Since the C$_{60}^-$ units in a polymer chain are 
rigidly linked in a same orientation, the C$_{60}^-$ chains produce
coherent electric field gradients which induce an anisotropic (quadrupolar)
deformation of the electron shell of the alkalis.
We model the corresponding charge distribution of each alkali by
a symmetric linear dumbbell
centered on lines along $\vec{a}$ directions.
On a same line, these induced dumbbells are parallel with
their axis perpendicular to $\vec{a}$
and a same orientation angle $\nu$ with the $\vec{c}$ axis.
We consider chains of alkali dumbbells, where the rigid chain
aspect is not imposed by intrachain interactions but by the
surrounding C$_{60}^-$ chains.
The orientational motion of an alkali dumbbell is characterized by
the SAF's $s_2(\vec{n})=\sin(2\nu(\vec{n}))$, where $\vec{n}$
is again a two dimensional array labeling the chains.
The quadrupole-quadrupole interaction between the C$_{60}^-$
chains and the surrounding alkalis is then given by
\begin{eqnarray} 
 H_{sS}=  \sum_{\vec{n} \vec{n}'}  
 (\lambda_b(\vec{n},\vec{n}')+\lambda_c(\vec{n},\vec{n}'))\, 
 S_2(\vec{n})\, s_2(\vec{n}') , \label{4}
\end{eqnarray}
where $\lambda_b(\vec{n},\vec{n}')$ accounts for the two
alkalis at $\pm a/2$ on the chains $X(\vec{n}')=\pm (\vec{r}_1-\vec{r}_2)/2$
and where $\lambda_c(\vec{n},\vec{n}')$ describe the interaction
with one alkali on the chains $X(\vec{n}')=\pm (\vec{r}_1+\vec{r}_2)/2$.
In Fourier space we have
\begin{mathletters} 
\begin{eqnarray} 
 & &H_{sS}=   \sum_{\vec{q}}
 \lambda(\vec{q})\, S_2(\vec{q})\, s_2(-\vec{q}) , \label{5a} \\
 & &\lambda(\vec{q})=4 \lambda_b\, \cos \left( \frac{q_y b}{2} \right) 
 +2 \lambda_c \cos \left( \frac{q_z c}{2} \right) . \label{5b}
\end{eqnarray}
\end{mathletters}
We take dumbbells with charges $q_A$ at
distances $\pm d_A$ from the center.
Here $A$ refers to K, Rb, or Cs. 
The numerical values of $d_A$ (Table~I) are the 
average radii of valence electron
$d$ shells calculated with atomic wave functions 
$3d_{3/2}$, $4d_{3/2}$, $5d_{3/2}$ for K$^+$, Rb$^+$ and Cs$^+$
respectively. 
We consider $d$-shells because they can support an
electric quadrupole moment.
We observe that the values for Cs$^+$
and Rb$^+$ are close to each other but differ from K$^+$.
The alkali ions are isoelectronic with the rare gas atoms Ar, Kr, Xe.
There the role of excited $d$ states has been found to be important
in the explanation of the face centered cubic structure \cite{Nie}.
With the same value $q_A=0.15 |e|$ for the three cases, we have
calculated the interaction energies $\lambda_b$ and $\lambda_c$
quoted in Table I. The quantity $| \lambda(\vec{q}) |$
is maximum for $\vec{q}=\vec{q}_{\Gamma}$
in contradistinction with $|J(\vec{q})|$.
The intra ionic restoring forces which oppose the deformation of the
electron shells of the
alkalis are described by a sum of single particle energy terms
\begin{eqnarray} 
 H_{ss}=g_A  \sum_{\vec{n}}  s_2^2(\vec{n})=
 g_A \sum_{\vec{q}} 
 s_2(\vec{q})\, s_2(-\vec{q}) , \label{6}
\end{eqnarray}
with $g_A>0$.
The self energy $g_A$ is inversely proportional to the
quadrupolar electronic polarizability and hence
$g_{Cs}<g_{Rb}<g_K$ (see Table I).
These concepts are inspired from the shell model of lattice 
dynamics \cite{Mig}.
The direct interchain coupling of alkali quadrupoles is numerically
found to be small and will be neglected.
We consider the total interaction $H=H_{SS}+H_{sS}+H_{ss}$.
The induced alkali quadrupoles follow adiabatically the motion of
the C$_{60}^-$ chains.
For a given configuration $\{S_2(\vec{q}) \}$ of the latter, we
minimize $H$ with respect to $s_2(\vec{q})$ and find
\begin{eqnarray} 
 s_2(\vec{q})=-\frac{1}{2} \frac{\lambda(\vec{q})}{g_A} S_2(\vec{q}), \label{7}
\end{eqnarray}
Substituting this result for $s_2$ in $H$ we get
\begin{mathletters} 
\begin{eqnarray} 
 H =   \frac{1}{2} \sum_{\vec{q}}
 \tilde{J}(\vec{q})\, S_2(\vec{q})\, S_2(-\vec{q}) , \label{8a} 
\end{eqnarray}
where
\begin{eqnarray} 
 & &\tilde{J}(\vec{q})=J(\vec{q})+C(\vec{q}) ,  \label{8b}  \\
 & &C(\vec{q})= -\frac{1}{2} 
 \frac{\lambda^2(\vec{q})}{g_A} . \label{8c}
\end{eqnarray}
\end{mathletters}
The coupling to the alkalis leads to an effective orientational
interaction $C(\vec{q})$ between C$_{60}^-$ chains.
This interaction is attractive and since
$|C(\vec{q})|$ with $\lambda(\vec{q})$ is maximum at 
$\vec{q}=\vec{q}_{\Gamma}$,
it favors a ``ferrorotational" structure (space group $I2/m$).
On the other hand the direct quadrupolar interaction 
between C$_{60}^-$ chains with
$|J(\vec{q}_Z)|>|J(\vec{q}_{\Gamma})|$ favors the antiferrorotational
structure. 
The system of interacting polymer chains described by Eq.~(\ref{8a})
chooses the lowest energy structure which is $Pmnn$ or $I2/m$
depending on whether $|\tilde{J}(\vec{q}_Z)|>|\tilde{J}(\vec{q}_{\Gamma})|$,
$i.e.$ $J_a>\lambda_b \lambda_c/g_A$,
or the opposite holds respectively.
Using the numerical values from Table I, we have plotted
$\tilde{J}(\vec{q})$ for $\vec{q}=(q_y=0,q_z)$ in Fig.~2.
For KC$_{60}$ \cite{Ste} with the small polarizability of the K$^+$
ion, the direct interchain interaction $J(\vec{q}_Z)$ dominates
and leads to $Pmnn$ while for RbC$_{60}$ \cite{Lau}
and even more for CsC$_{60}$ \cite{Rou}
the alkali mediated interaction $C(\vec{q}_{\Gamma})$
dominates and leads to $I2/m$.
%
\begin{figure} 
\vspace{-1.4cm}
\centerline{ 
\epsfig{file=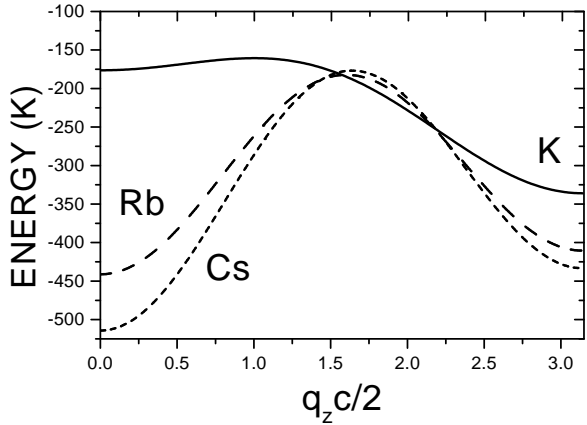,width=0.47\textwidth} 
} 
\vspace{-4.6cm}
\caption{
Interchain energy $\tilde{J}(\vec{q})$, units K.
} 
\label{fig2} 
\end{figure} 
A condensation of $S_2(\vec{q})$ at the BZ center leads, 
via coupling to the center of mass displacements
of the alkali ions (bilinear RT-coupling) to
$\epsilon_{yz}$ shear modes and hence to a deviation of the
$(\vec{b},\vec{c})$ angle $\alpha$ from 90$^0$
in the $I2/m$ structure.
However this RT-coupling is not the
driving process of the monoclinic structure.
Indeed, the deviations of $\alpha$ are very small \cite{Lau,Rou}.

In conclusion we have shown that the cycloaddition
between C$_{60}^-$ units leads to an orthorhombic
lattice structure.
The concomitant symmetry reduction produces a quadrupolar  electric
charge distribution on the C$_{60}^-$ units.
The interaction between C$_{60}^-$ chains has two competing components:
a direct quadrupole-quadrupole interaction $J(\vec{q})$
and an indirect one $C(\vec{q})$.
The latter is mediated by the induced quadrupoles on
the polarized alkali metal ions.
The direct quadrupolar interaction drives the
antiferrorotational structure $Pmnn$ while the indirect
one yields the ferrorotational structure $I2/m$.
In RbC$_{60}$ and CsC$_{60}$ with larger electronic
polarizability and quadrupolar radius $d_A$ of the alkalis,
$\lambda_b \lambda_c/g_A > J_a$, 
the ferrorotational structure is realized while 
in KC$_{60}$, $J_a > \lambda_b \lambda_c/g_A$,  
the antiferrorotational structure is realized.
Within the present theory, the alkalis play a specific
role beyond the function of lattice spacers.
The study of alkali specific effects is a problem of broad
interest and likely to be relevant for the
understanding of superconducting fullerides \cite{Ceg}.

We acknowledge useful discussions with K. Knorr, P. Launois
and R. Moret.
This work has been financially supported by
the Fonds voor Wetenschappelijk Onderzoek, Vlaanderen.


\vspace{0.5cm}

\end{document}